\documentclass[pra,twocolumn,superscriptaddress,nobibnotes]{revtex4}

\usepackage{graphicx}% Include figure files
\usepackage{dcolumn}% Align table columns on decimal point
\usepackage{bm}% bold math
\usepackage{times,mathptmx}
\usepackage{color}
\usepackage{hyperref}
\usepackage{placeins}

\begin{document}

\title{A chip-scale integrated cavity-electro-optomechanics platform}

%\author{Martin Winger~,$^1$ Tim D.\ Blasius,~$^1$ Thiago P. Mayer Alegre,~$^1$ Amir H. Safavi-Naeini,~$^1$ Se\'an Meenehan,~$^1$ Justin Cohen,~$^1$ S\o ren Stobbe,~$^{2,3}$ and Oskar Painter~$^1$}

\author{Martin Winger}
\email{winger@caltech.edu}
\author{Tim Blasius}
\affiliation{Thomas J.\ Watson, Sr., Laboratory of Applied Physics, California Institute of Technology, Pasadena, CA 91125, USA}
\author{Thiago P.\ Mayer Alegre}
\affiliation{Thomas J.\ Watson, Sr., Laboratory of Applied Physics, California Institute of Technology, Pasadena, CA 91125, USA}
\affiliation{Current address: Instituto de F\'isica ``Gleb Wataghin'', Universidade Estadual de Campinas, UNICAMP 13083-859, Campinas, SP, Brazil}
\author{Amir H.\ Safavi-Naeini}
\author{Se\'an Meenehan}
\author{Justin Cohen}
\affiliation{Thomas J.\ Watson, Sr., Laboratory of Applied Physics, California Institute of Technology, Pasadena, CA 91125, USA}
\author{S\o ren Stobbe}
\affiliation{DTU Fotonik, Department of Photonics Engineering, Technical University of Denmark, \O rsteds Plads 343, DK-2800 Kgs.\ Lyngby, Denmark}
\affiliation{Current address: Niels Bohr Institute, University of Copenhagen, Blegdamsvej 17, DK-2100 Copenhagen, Denmark}
\author{Oskar Painter}
\email{opainter@caltech.edu}
%\homepage{http://copilot.caltech.edu}
\affiliation{Thomas J.\ Watson, Sr., Laboratory of Applied Physics, California Institute of Technology, Pasadena, CA 91125, USA}

%\address{$^1$Thomas J.\ Watson, Sr., Laboratory of Applied Physics, California Institute of Technology, Pasadena, CA 91125, USA\\
%$^2$DTU Fotonik, Department of Photonics Engineering, Technical University of Denmark, \O rsteds Plads 343, DK-2800 Kgs.\ Lyngby, Denmark\\
%$^3$Current address: Niels Bohr Institute, University of Copenhagen, Blegdamsvej 17, DK-2100 Copenhagen, Denmark}

%\email{winger@caltech.edu}

%\email{opainter@caltech.edu}

%\homepage{http://copilot.caltech.edu}

\begin{abstract}
We present an integrated optomechanical and electromechanical nanocavity, in which a common mechanical degree of freedom is coupled to an ultrahigh-$Q$ photonic crystal defect cavity and an electrical circuit. The system allows for wide-range, fast electrical tuning of the optical nanocavity resonances, and for electrical control of optical radiation pressure back-action effects such as mechanical amplification (phonon lasing), cooling, and stiffening. These sort of integrated devices offer a new means to efficiently interconvert weak microwave and optical signals, and are expected to pave the way for a new class of micro-sensors utilizing optomechanical back-action for thermal noise reduction and low-noise optical read-out.%\vspace{1cm}
\end{abstract}

\maketitle

%\ocis{(230.5298)~Photonic crystals; (230.3120)~Integrated optics devices; (230.4110)~Modulators; (230.4685)~Optical microelectromechanical devices; (220.4880)~Optomechanics; (280.4788)~Optical sensing and sensors; (350.2460)~Filters, interference; (350.4238)~Nanophotonics and photonic crystals}

%%%%%%%%%%%%%%%%%%%%%%% References %%%%%%%%%%%%%%%%%%%%%%%%%

%%%%%%%%%%%%%%%%%%%%%%%%%%  body  %%%%%%%%%%%%%%%%%%%%%%%%%%
\section{Introduction}

The usually feeble force associated with radiation pressure \cite{Nichols01}, a manifestation of the mechanical momentum carried by all electromagnetic waves, has recently proven to be quite effective in manipulating and detecting the motion of micro- and nanomechanical objects embedded within a resonant cavity \cite{Kippenberg07,Kippenberg08,Favero09}. The simplest form of a cavity-mechanical system consists of a resonant electromagnetic cavity with its resonance frequency dispersively coupled to the position of a mechanical object. In such a cavity-based scheme, a narrowband electromagnetic source is used to pump the cavity.  Mechanical motion translates into modulation in the stored intra-cavity electromagnetic field, and through the filtering properties of the cavity, results in an imprinting of the mechanical motion on the electromagnetic signal.  The resonant enhancement of the pump's radiation pressure in turn, yields strong back-action effects which modify the dynamic mechanical and optical properties of the coupled system.  Dynamical back-action effects can include optical stiffening of the mechanical structure \cite{Favero09,Braginskii77,Braginskii92,Rosenberg09,Eichenfield09a}, damping or amplification of the mechanical motion \cite{Braginskii92,Arcizet06,Kippenberg05,Gigan06}, or electromagnetically induced transparency \cite{Weis10,Teufel11_strong_coupling,Safavi11_EIT}. 

Cavity-mechanical systems demonstrating near quantum-limited position read-out and strong radiation pressure back-action have been realized both in the optical \cite{Schliesser08,Chan11_arXiv} and the microwave frequency domains \cite{Regal08,Teufel11}. In the optical domain one has the advantage of shot-noise limited read-out (even at room temperature) and large radiation pressure coupling due to the relatively large operating frequency, whereas in the microwave domain one has the distinct benefit of simple electrical interfacing and compact, robust packaging.  Here we present a chip-scale platform for integrating cavity-optomechanics with conventional micro-electromechanical systems (MEMS) in which the mechanical degree of freedom is strongly coupled via radiation pressure to both an electrical circuit as well as a high-$Q$ optical cavity \cite{Lee10}.  Using an integrated photonic crystal device we demonstrate wide-band ($\sim$19~nm) electromechanical tuning of the optical cavity resonance, near shot-noise-limited optical read-out of mechanical motion, and electromechanical locking of the optical cavity to a fixed laser source.  By combining these device attributes, a series of key optomechanical back-action effects are also realized, including optical stiffening, back-action cooling, and phonon lasing.  It is envisioned that these coupled electro- and optomechanical systems, driven by radiation pressure and packaged in a chip-scale form factor, may be used to create sensors of electrical \cite{Safavi11}, force \cite{Schliesser08,Regal08}, acceleration, or mass \cite{Ekinci05} operating at the quantum limits of sensitivity and bandwidth.

\section{A tunable slotted-waveguide photonic crystal cavity}

As discussed above, in this work we seek to develop a common platform for cavity electro- and optomechanics, in which both electrical and optical signals are coupled to a common mechanical degree of freedom \cite{Lee10}.  Planar photonic crystals (PCs) are particularly promising to this end, since they provide the potential for on-chip integration with well-established microwave and micro-electromechanical systems (MEMS) technologies, and large radiation pressure coupling due to their nanoscale optical mode volumes \cite{Eichenfield09a,Eichenfield09b,Safavi10}.  Electromechanical control of microcavities has been shown previously in one-dimensional zipper and double-membrane cavities \cite{Frank10,Perahia10,Midolo11}. These approaches, however, were either limited by low tuning speed, high leakage-currents, or the use of low-$Q$ cavities, which prohibited the observation of radiation back-action effects. 

\begin{figure*}[ht]
\centering\includegraphics[width=1.8\columnwidth]{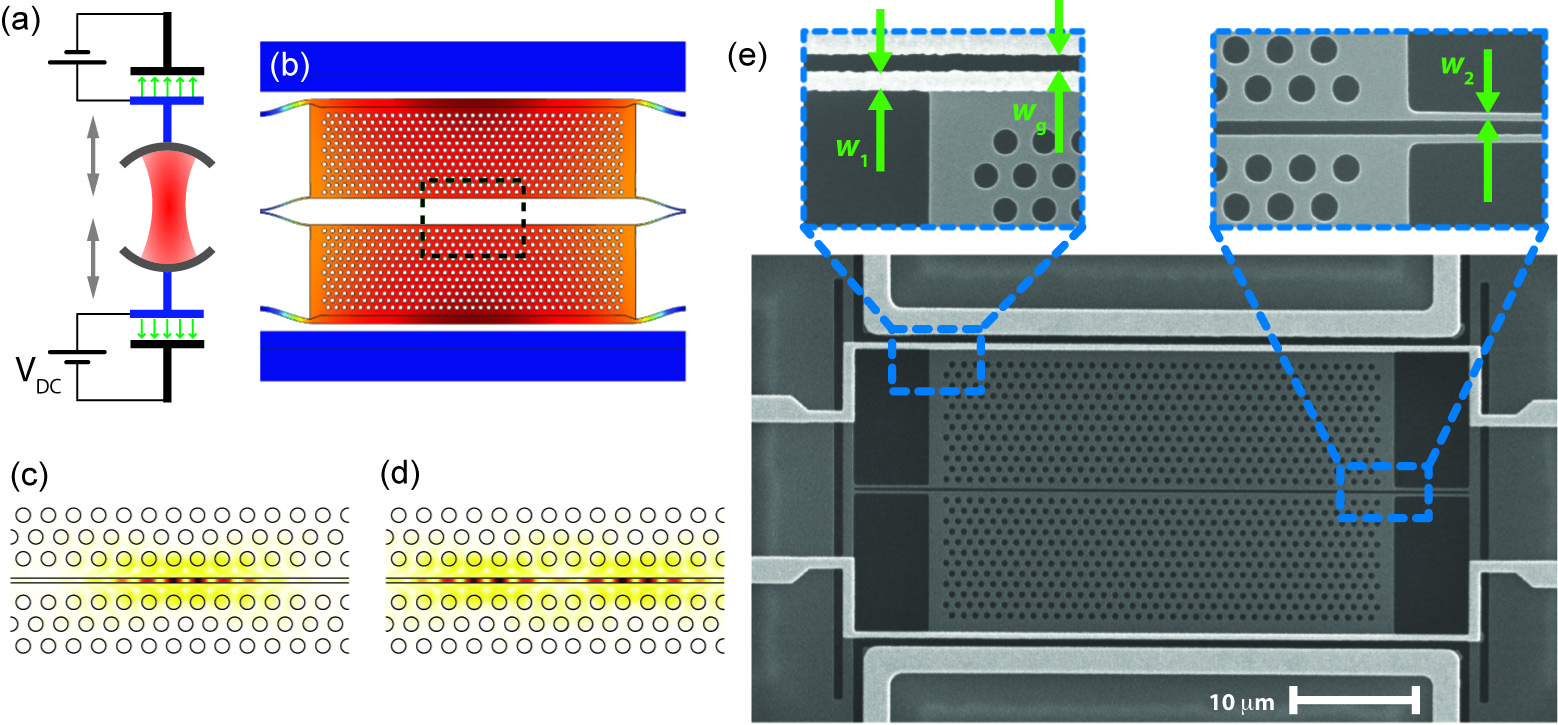}
\caption{\label{Fig1}
	(a) Scanning Fabry-P\'erot cavity example of an electro-optomechanical system, in which cavity mirrors are attached to capacitive actuators.
	(b) Displacement profile of the PC implementation of an electro-optomechanical cavity. The cavity is formed as a waveguide defect in between two individual PC membrane halves (region outlined by the black rectangle), the distance between which can be adjusted using an electrostatic force generated between pairs of metal wires.
	(c) \& (d) Electric field distribution $|\mathbf{E}|^2$ of the first ((c)) and second order (d)) optical cavity modes.
	(e) Scanning-electron micrograph of a processed device in a double-capacitor configuration. The PC membrane is suspended on struts with $w_1 = 250\ \mathrm{nm}$ and $w_2 = 80\ \mathrm{nm}$, a zoom-in of which is shown in (d).}
\end{figure*}

The photonic crystal structure studied in this work, along with the conceptually similar tunable Fabry-Perot cavity, are sketched in Figs.~\ref{Fig1}(a) and (b).  The PC structure is based on a silicon membrane slotted cavity, in which the optical cavity field is localized to the air slot between two PC membranes that are suspended on flexible struts. Two pairs of metal contacts on each membrane act as capacitive MEMS actuators that provide for electromechanical control of membrane motion and the cavity air slot width. Localization of light in the optical cavity is determined by a two-step compression of the PC lattice constant \cite{Song05} along the length of the slotted PC waveguide formed from the two PC membrane halves \cite{Safavi10}. In this work, the PC structure was fabricated with a lattice constant of $a=470\ \mathrm{nm}$, a relative hole radius of $r/a = 0.285$, and a slot width of $s/a = 0.21\ \mathrm{nm}$ so as to produce cavity modes in a wavelength band around $1500$~nm with high-$Q$ and large radiation pressure coupling. Theoretical estimates for the optical cavity mode frequencies and radiation rates were calculated using a finite-elements method (FEM) solver which is part of the COMSOL Multiphysics \cite{COMSOL} software package. The cavity is found to support two high-$Q$ modes (theoretical $Q>10^6$), FEM simulations of which are shown in Fig.~\ref{Fig1}(c) and (d).

The strong light confinement in the slot region makes the optical mode frequency ($\omega_c$) highly sensitive to the separation $s$ of the two membranes with a theoretical optomechanical coupling $g'_{\mathrm{OM}} = \partial \omega_\mathrm{c} / \partial s = \omega_\mathrm{c}/L_\mathrm{OM} = 2\pi \times 152\ \mathrm{GHz/nm}$ obtained from FEM simulations. The electrostatic actuators are formed by pairs of gold contacts that together with the underlying silicon form capacitors (capacitance $C$) which create an electrostatic force $F_\mathrm{el}=(1/2)(\mathrm{d}C/\mathrm{d}w_g)V_\mathrm{a}^2$ when applying a voltage $V_\mathrm{a}$ across the capacitor gap $w_g$ \cite{Alegre10}. $F_\mathrm{el}$ leads to contraction of the capacitors, thus increasing $s$ and leading to a blue-shift of the cavity resonances. Figure~1(e) shows a scanning-electron micrograph of a device fabricated on a microelectronics SOI wafer. The cavity membranes are suspended on $l=3\ \mathrm{\mu m}$ long struts of width $w_1 = 250\ \mathrm{nm}$ and $w_2 = 80-150\ \mathrm{nm}$, respectively, yielding estimated effective spring constants for in-plane motion on the order of $k_\mathrm{eff} \approx 50\ \mathrm{N/m}$. For a metal layer thickness of 200~nm and capacitor gaps of $w_g = 200-250\ \mathrm{nm}$ we estimate $C \sim 0.7\ \mathrm{fF}$ and $F_\mathrm{el} \sim 1.5\ \mathrm{nN/V}^2$.

\subsection{Fabrication}
Samples were fabricated from silicon-on-insulator material from SOITEC. A lift-off mask for the metal contacts is defined by electron-beam lithography in ZEP-520A positive e-beam resist. We then deposit a 5nm/200nm thick Cr/Au layer in an electron-beam evaporator and strip the resist with the excess metal on top in Microposit 1165 photoresist remover. A fresh layer of ZEP-520A is applied, and the etch-pattern for the PC structures, together with the necessary cut-outs for capacitor gaps, membrane suspensions, and strain-relief slices is exposed. The pattern is transferred into the silicon by a radio-frequency plasma of $\mathrm{C}_4\mathrm{F}_8/\mathrm{SF}_6$ chemistry. The excess e-beam resist is removed by cascaded immersion into trichloroethylene, Microposit 1165 remover, and a 10~min etch in Piranha solution (3:1 $\mathrm{H}_2\mathrm{SO}_4$:$\mathrm{H}_2\mathrm{O}_2$) at $120^\circ \mathrm{C}$. The cavity membranes are released from the underlying $\mathrm{SiO}_2$ layer by immersion into $48\%$ Hydrofluoric acid. Cleaning of the sample surface is finalized by an additional Piranha cleaning step, followed by a rinse in de-ionized water and a 1~min immersion into 1:10 $\mathrm{HF:H}_2\mathrm{O}$. Finally, samples are glued to a copper sample holder using GE varnish and electrically contacted with gold wires by ultrasonic wire-bonding.

\section{Optical and mechanical characterization}

\subsection{Optical spectroscopy}

\setcounter{totalnumber}{2}
\begin{figure}[Ht]
\centering\includegraphics[width=0.9\columnwidth]{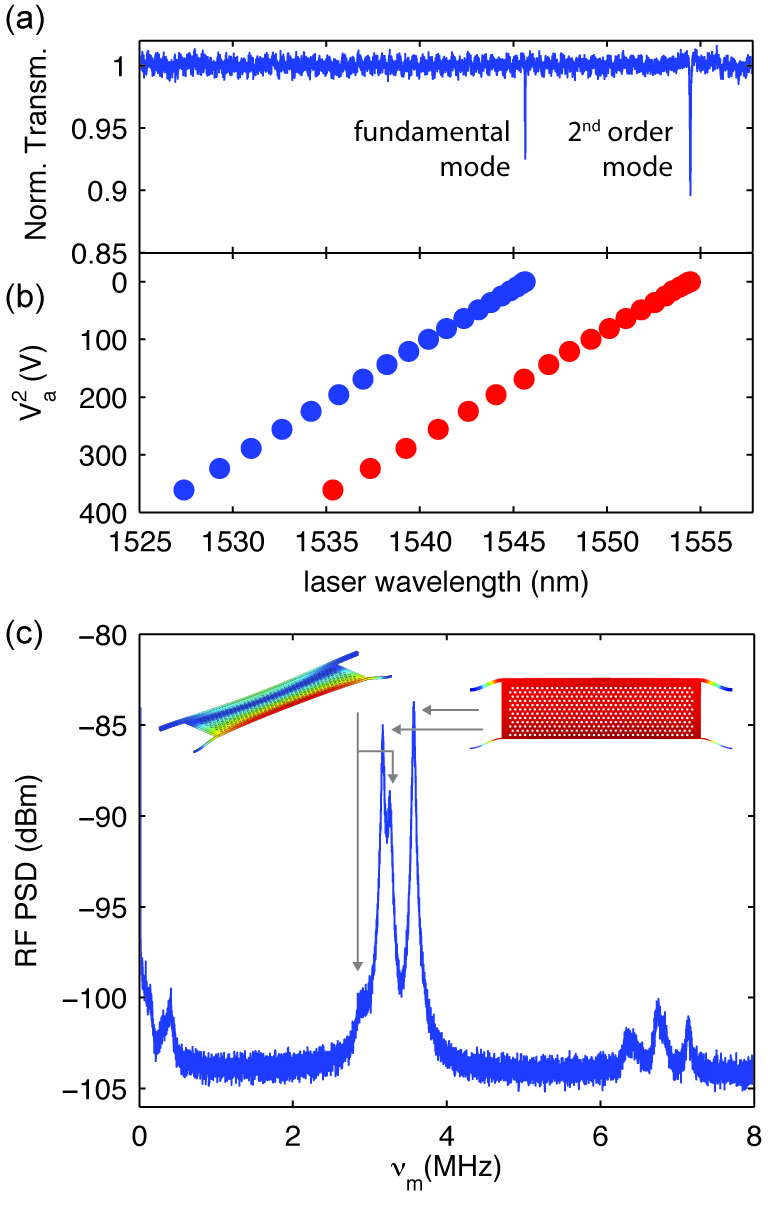}
\caption{(a) Plot of the normalized transmission spectrum of a device with zero applied voltage showing both the fundamental and the second-order optical resonance. 
	(b) Cavity resonance wavelengths from (a) versus applied voltage ($V_\mathrm{a}^2$), indicating quadratic wavelength-tuning of the cavity modes with tuning parameter $\alpha = 0.051\ \mathrm{nm/V^2}$.
	(c) RF power spectral density of laser light transmitted through the second order cavity mode. The resonances at 3.18, 3.28, and 3.61~MHz correspond to modes with hybridized in- and out-of-plane character. The insets show FEM-simulations of the eigenmodes of a single membrane half in top- and sideview.}\label{Fig2}
\end{figure}
The PC cavities are optically investigated by resonant transmission spectroscopy using a near-field technique based on a dimpled tapered optical fiber, the long-range evanescent field of which is brought into optical contact with the cavity \cite{Michael07}. A swept-wavelength narrow-band telecommunications test laser allows for obtaining cavity transmission spectra. Figure~\ref{Fig2}(a) shows the transmission spectrum of a device with $w_2=150\ \mathrm{nm}$. The two resonances at $1545.63$~nm and $1554.45$~nm correspond to the cavity modes depicted in Fig.~1(c) and (d), respectively. When increasing the applied voltage $V_\mathrm{a}$, these resonances blue-shift, as can be seen in Fig.~\ref{Fig2}(b). For a maximum applied voltage of $V_\mathrm{a} = 19\ \mathrm{V}$ the fundamental (second order) mode reaches a total shift of {$-18.3$~nm} ($-19.1$~nm) or {$+2.32$~THz} ($+2.4$~THz) without a noticeable reduction of the optical $Q$-factor. As expected, cavity tuning follows a quadratic voltage dependence. Defining the tunability $\alpha$ by $\Delta\lambda_\mathrm{c} = \alpha\cdot V_\mathrm{a}^2$, this corresponds to a measured $\alpha = 0.051\ \mathrm{nm}/\mathrm{V^2}$, in good correspondence with the FEM electromechanical simulations of the structure. For devices with $w_2 = 80\ \mathrm{nm}$, we were able to achieve tunabilities up to $\alpha = 0.088\ \mathrm{nm}/\mathrm{V^2}$ (see Appendix~\ref{SecStudyFabDev}). The accessible tuning range of a given device is limited by electrical arching between the contacts, which occurs around $V_\mathrm{max} \approx 20\ \mathrm{V}$ in a Nitrogen atmosphere at ambient pressure. Also, due to the large parallel resistance in excess of $400\ \mathrm{G\Omega}$, current flow in these structures is negligible, minimizing heating and allowing for ultralow power operation. Viewed as a wide-range ($> 2$~THz) tunable optical filter (bandwidth $\sim 1$~GHz) operating in the telecom C-band, or as a narrowband modulator/switch with ultra-low switching voltage ($V_\pi = 10\ \mathrm{mV}$ at a bias voltage of $V_\mathrm{a}=10$~V), the present device performance is impressive due in large part to the strong optomechanical coupling. 

\begin{figure*}[t]
\centering\includegraphics[width=1.8\columnwidth]{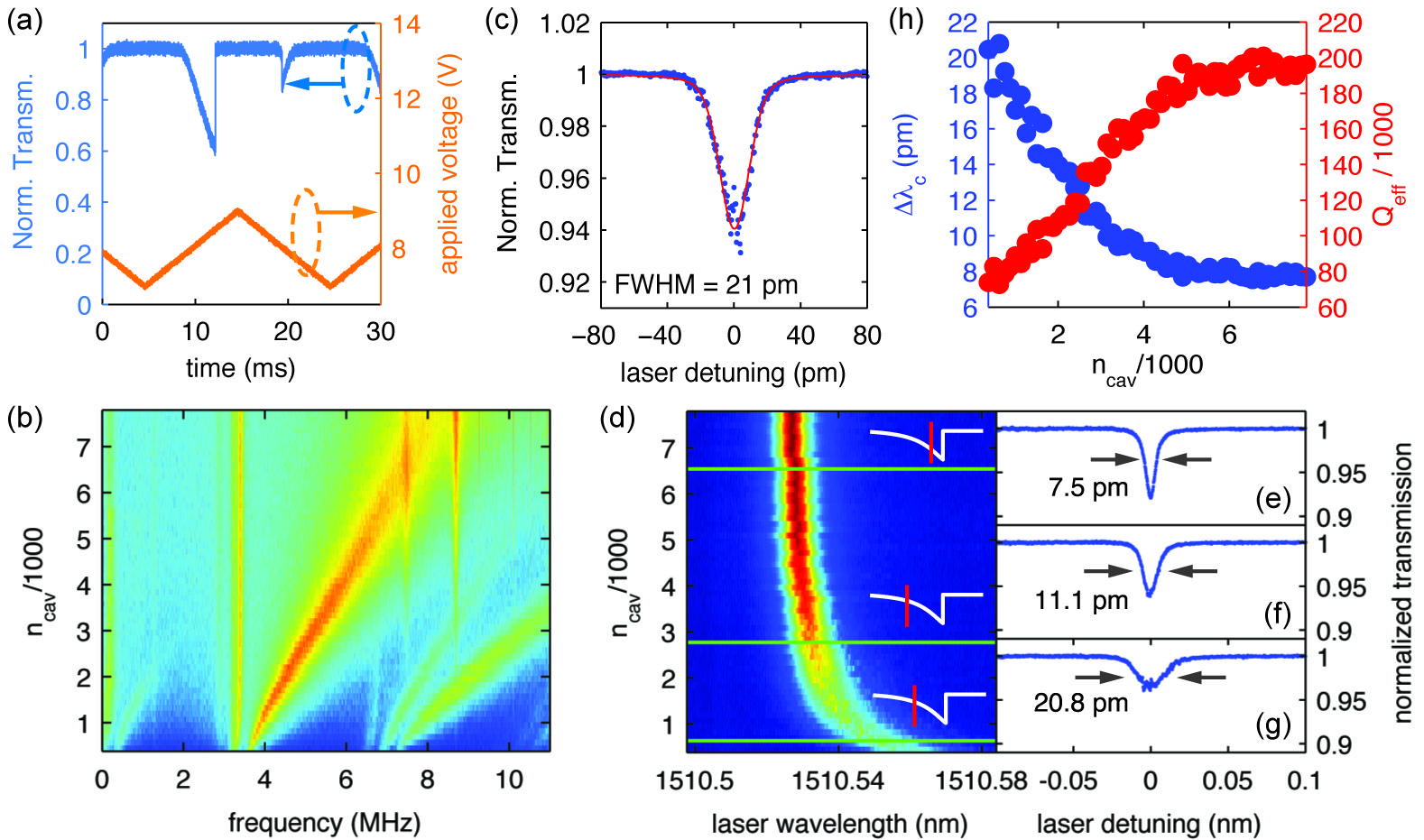}
\caption{Optical spring reduction of thermal noise:
	(a) Transmission spectrum (blue) of the second order cavity mode. The actuators are driven by a triangular signal (orange) with an amplitude of 1~V and a frequency of 50~Hz. 
	(b) RF spectra of the pump laser transmission as function of $n_\mathrm{cav}$. The optical spring effect shifts the mechanical mode at $3.6$~MHz to higher frequencies.
	(c) Normalized transmission spectrum of the fundamental cavity mode (blue) of a device with $\alpha = 0.055\ \mathrm{nm/V^2}$ together with a Voigt fit (red).
	(d) False color plots of the transmission scans of the fundamental cavity mode as function of the intracavity photon number $n_\mathrm{cav}$ in the second order mode.  The horizontal green lines indicate the intra-cavity photon numbers at which the individual scans in (e)--(g) were taken.
	(h) Linewidth of the fundamental cavity mode as function of $n_\mathrm{cav}$ in the second order mode. Blue dots show the FWHM linewidth extracted from (d) while the red dots show the linewidth as an effective optical $Q$-factor.
	}\label{Fig3}
\end{figure*}

\subsection{Mechanical mode spectrocsopy}
In addition to the optical properties of the PC cavity, the mechanical mode structure of the presented system can be investigated by monitoring the radio-frequency (RF) power spectral density (PSD) of laser light transmitted through a cavity mode. To this end, we launch a tunable external cavity diode laser into the fundamental or second order cavity mode, and actively stabilize the cavity frequency to a detuning of half an optical linewidth from the laser. The transmitted pump light is detected on a high-speed photodetector (125~MHz bandwidth), and the fluctuation power spectral density of the photocurrent is computed with a high-speed digitizing oscilloscope. In Fig.~\ref{Fig2}(c) this is shown for transmission of a probe laser through the second order cavity mode. The strongly transduced resonant features between $3$~MHz and $3.6$~MHz correspond to mechanical modes of the structure that originate from the in-plane tuning mode (right inset in Fig.~2(c)) of the two individual membranes, split by fabrication asymmetries. Moreover, hybridization with a near-resonant out-of-plane (flexural) mode (see left inset in Fig.~2(c)) that originates from the breaking of out-of-plane symmetry induced by the presence of the top metal contacts gives rise to the additional features at 2.9 and 3.15~MHz (see Appendix~\ref{SecMechSpectra}). The mechanical modes shown here can also be resonantly addressed by driving the actuators with a sinusoidal modulation voltage (see Appendix~\ref{SecAcTuning}). The $Q$-factors of the mechanical modes were found to be in the range 50--100, limited by air-damping~ \cite{Eichenfield09a}, thus allowing for high-speed tuning of the structure at rates limited by the mechanical time-constant of $20\ \mathrm{\mu s}$.

Despite its unique benefits for the readout and manipulation of micromechanical motion, optomechanical back-action has hitherto not found technological application in large part due to the need for elaborate tunable laser-sources to control the relative cavity-pump laser detuning. In the presented system, however, frequency tunability is solely afforded by electromechanical actuation, thus allowing for the study of optomechanical effects using simple fixed-frequency laser sources. As an example of this, Fig.~\ref{Fig3}(a) shows an oscilloscope trace of the transmission of a strong pump laser ($P_\mathrm{i} = 270\ \mathrm{\mu W}$) through the second order cavity mode (blue curve) while applying a 50~Hz triangular wave to the actuators (orange curve). Both the triangular shape of the transmission curve and the asymmetry between forward- and backward scans arise from the well-known thermal bistability of silicon microcavities \cite{Barclay05}. Using electromechanical frequency tuning then, we can actively lock the cavity to a fixed-frequency pump laser. To this end, we actively control the cavity electrical contacts with a commercial PI-control loop. The error signal is obtained from the transmission level of the pump laser which therefore is proportional to the intracavity photon number $n_\mathrm{cav}$. Additionally, although not performed here, using a feedback loop of sufficient bandwidth allows for active feedback cooling (``cold-damping'') and amplification of the mechanical mode \cite{Kleckner06,Lee10}.

\section{Optical stiffening}

As an example of electrically controlled optomechanical back-action, we study the optical spring effect by tuning the cavity in resonance with the blue-detuned pump laser. Figure~\ref{Fig3}(b) shows a series of RF-modulation spectra while changing the intracavity photon number $n_\mathrm{cav}$ in the second order cavity mode. This is achieved by actively locking the tunable cavity to different levels of the pump laser transmission as described above. The higher frequency mode initially at $3.61$~MHz is renormalized by the radiation pressure coupling to the internal cavity field into the in-plane differential mode of Fig.~\ref{Fig1}(b)~\cite{Rosenberg09}, and shifts to $\approx 8\ \mathrm{MHz}$ for $n_\mathrm{cav} = 7\,500$ (note that the lower frequency mode at $3.3$~MHz shifts very little, as it is renormalized to the uncoupled common mode of motion between the membrane halves). The observed frequency shift is consistent with $g_\mathrm{OM} = 2\pi\times 215\ \mathrm{GHz/nm}$, in reasonable agreement with the theoretically expected value.

\begin{figure*}[ht]
\centering\includegraphics[width=1.8\columnwidth]{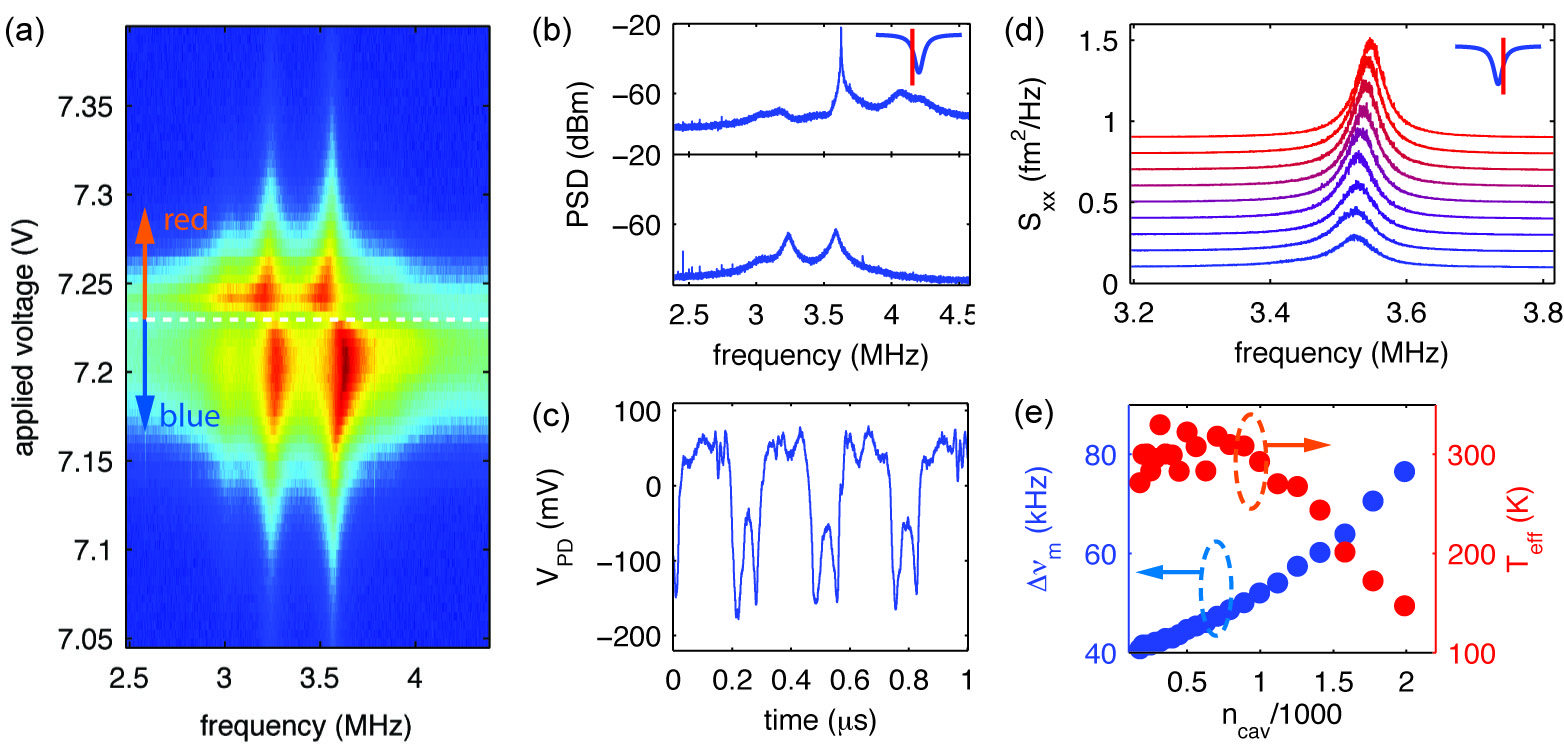}
\caption{Electrically controlled optomechanical back-action:
	(a) False-color plot of RF optical transmission spectra as a function of $V_\mathrm{a}$. For blue detuning we observe stiffening and amplification of mechanical motion, whereas we observe softening and damping for red detuning. 
	(b) RF spectra for a blue-detuned pump laser below (lower) and above the lasing threshold (upper panel). 
	(c) Time trace of the cavity optical transmission in the phonon lasing regime.
	(d) Waterfall plot of the RF optical transmission spectra of the mechanical modes in the cooling regime, with the pump-laser a half-linewidth red-detuned from the cavity.  Curves go from red to blue as $n_{cav}$ is increased.
	(e) Plot of the higher-frequency $3.6$~MHz mechanical mode linewidth (blue dots) and effective temperature (red dots) versus $n_{cav}$ under red-detuned pumping.  
	}\label{Fig4}
\end{figure*}
The optical spring effect is rather unique in that it only affects the dynamic spring constant of the mechanical system responding to fluctuations around mechanical equilibrium, but leaves alone the static stiffness of the structure \cite{Rosenberg09}.  Increasing the wide-range tunability of a micro-mechanical device by reducing the spring constant $k_\mathrm{eff} = m_\mathrm{eff} \omega_m^2$ naturally leads to a compromise in which the noise is increased due to thermal processes.  The frequency jitter of the cavity resonance in the highly flexible structures of this work is estimated to be $\Delta\lambda_\mathrm{rms} = (\lambda_c/L_{\mathrm{OM}}) \sqrt{k_\mathrm{B}T/m_\mathrm{eff}\omega_m^2} = 18.1\ \mathrm{pm}$, almost the entire measured optical linewidth of $22$~pm. As a result, time-averaged transmission scans like that of the fundamental optical cavity resonance shown in Fig.~\ref{Fig3}(c) are predominantly thermally broadened. The red line here shows a fit assuming a Voigt line profile (see Appendix~\ref{SecVoigtFits}) that allows us to estimate an intrinsic linewidth of $6$--$9$~pm with thermal Gaussian line broadening of $\approx18$~pm.  Using optomechanical back-action this thermal noise can be overcome, without sacrifice in tunability, by increasing $k_\mathrm{eff}$ using the optical spring effect.

In order to investigate the effect of the reduction of thermal membrane motion by increasing $k_\mathrm{eff}$ via the optomechanical spring effect, we monitor transmission spectra of the fundamental cavity mode as function of $n_\mathrm{cav}$ stored in the second order (pump) cavity mode. To this end, we use two separate telecom external-cavity diode lasers that are combined via a fiber-based optical beamsplitter before entering the fiber taper and that are individually detected after being separated by a fiber-based transmission/reflection bandpass filter at the taper output. One laser (pump laser) is kept at a fixed wavelength close to the second order optical cavity mode. Again, the detuning with the pump laser can then be controlled electrostatically. At the same time, the second laser is swept across the fundamental mode, resulting in the transmission spectra shown in Fig.~\ref{Fig3}(d) for various values of $n_\mathrm{cav}$. As $n_\mathrm{cav}$ increases, the cavity modes red-shift due to heating of the structure, which counter-acts the electrostatically induced blue-shift and results in the saturation of cavity tuning. At the same time, the linewidth of the fundamental cavity mode decreases significantly, as can be seen from the cuts through Fig.~\ref{Fig3}(d) shown in Figs.~\ref{Fig3}(e)--(g). The linewidths $\Delta\lambda_\mathrm{c}$ extracted from the transmission curves are shown as the blue bullets in Fig.~\ref{Fig3}(h), while the red bullets express the width as an effective $Q$-factor $Q_\mathrm{eff} = \lambda_\mathrm{c}/\Delta\lambda_\mathrm{c}$. While the initial linewidth is 21~pm, for $n_\mathrm{cav} = 7\,500$ we observe narrowing to 8~pm, corresponding to an intrinsic optical $Q$-factor of $200\,000$ (see SI for details). This is more consistent with the observed cavity linewidths of $\approx 3\ \mathrm{pm}$ ($Q\approx5\cdot10^5$) on nominally identical, mechanically rigid test cavities.

\section{Electrically controlled radiation-pressure back-action}

Using electromechanical control of the cavity frequency, we can also realize parametric amplification (phonon lasing) and back-action cooling. Figure~\ref{Fig4}(a) shows mechanical spectra of a different device while sweeping the fundamental cavity mode across resonance with a pump laser with $P_\mathrm{i} = 25\ \mathrm{\mu W}$. For a blue-detuned pump laser ($V_\mathrm{a}<7.23\ \mathrm{V}$) we observe stiffening of the mechanical modes -- similar to Fig.~\ref{Fig3}(h) -- while for red detuning we observe softening, indicated by a reduction of the mechanical mode frequencies. In the electro-optomechanical PC cavity, we can switch between the two regimes by using a fixed-frequency pump and simply changing a voltage.

Tuning the cavity such that the pump laser is blue detuned from the optical mode results in phonon lasing, while red detuning leads to cooling. In this system we can realize both regimes, as we will show in the following. Figure~\ref{Fig4}(b) shows RF-spectra for driving the system with $P_\mathrm{i} = 250\ \mathrm{\mu W}$ on the blue side, with detunings below (lower panel) and above (upper panel) the lasing threshold. Above threshold we observe line-narrowing and an enhancement of the mechanical resonance peak by approximately four orders of magnitude. In the time domain, this corresponds to a large, periodic modulation of the cavity transmission signal, as is evident from the time trace shown in Fig.~\ref{Fig4}(c). For red detuning, we observe cooling of the membrane motion. To this end, we mechanically anchor the fiber taper on one of the cavity membranes in order to suppress spurious out-of-plane modes  (see Appendix~\ref{SecAppBackaction}) and to reduce temporal drift of the fiber taper. We then lock the cavity a half-linewidth red from the pump laser and monitor RF-spectra while increasing the power $P_\mathrm{i}$ launched into the cavity. Figure~\ref{Fig4}(d) shows the membrane displacement spectral density $S_{xx}$ for a series of $n_\mathrm{cav}$. The resonance at 3.55~MHz corresponds to motion of a single membrane. Clearly, we observe optical damping of the mechanical mode with increasing photon number, evident from the increasing linewidth $\Delta\nu_\mathrm{m}$ of the resonance (blue dots in Fig.~\ref{Fig4}(e)). Measuring the area under the mechanical resonance (proportional to the phonon occupancy and thus the effective temperature) shows that in addition to optically-induced damping, there is optical cooling of the mechanical motion, with the effective temperature of the mode (red dots in Fig.~\ref{Fig4}(e)) reducing to $T_\mathrm{eff} \approx 150\ \mathrm{K}$.

\section{Conclusion}

As the above measurements indicate, the electro-optomechanical PC cavity structure demonstrated here provides a unique level of control of the optical properties of the device via electromechanical actuation, and, importantly, maintains the device characteristics necessary for strong radiation pressure coupling and optomechanical back-action.  This device architecture is also adaptable to integration with higher frequency (GHz) mechanical resonances (in the form of optomechanical crystals for instance \cite{Eichenfield09b}), enabling operation in the sideband-resolved regime critical for quantum applications \cite{Safavi11}.  Moreover, the scalability afforded by the CMOS-compatible, on-chip architecture of this device platform, allows for mass production, integration with on-chip electronics, and the straightforward engineering of coupled-cavity arrays.  Applications for such devices range from wavelength filters and ultralow-power optical modulators, to quantum-limited force sensors~\cite{Regal08,Schliesser08}.  Moreover, future devices in which an optical cavity and a radio-frequency resonant circuit share a common mechanical degree of freedom are foreseeable, paving the way for experiments at the interface of cavity opto- and electro-mechanics \cite{Lee10}, such as the implementation of efficient inter-band signal conversion \cite{Safavi11}.

\begin{acknowledgments}
This work was supported by the DARPA/MTO ORCHID program through a grant from the AFOSR, the DARPA/MTO MESO program through a grant from SPAWAR, and the NSF (CIAN grant no. EEC-0812072 through University of Arizona). S.St.\ gratefully acknowledges The Danish Council for Independent Research (Project No.\ FTP 10-080853).
\end{acknowledgments}

\onecolumngrid

\clearpage

\twocolumngrid

\renewcommand{\thefigure}{A\arabic{figure}}
\renewcommand{\theequation}{A\arabic{equation}}
\setcounter{figure}{0}
\setcounter{equation}{0}

\appendix
\section*{Appendix}

\section{Theory of capacitive cavity tuning}\label{SecTheory}
\begin{figure*}[htb]
\centering\includegraphics[width=1.8\columnwidth]{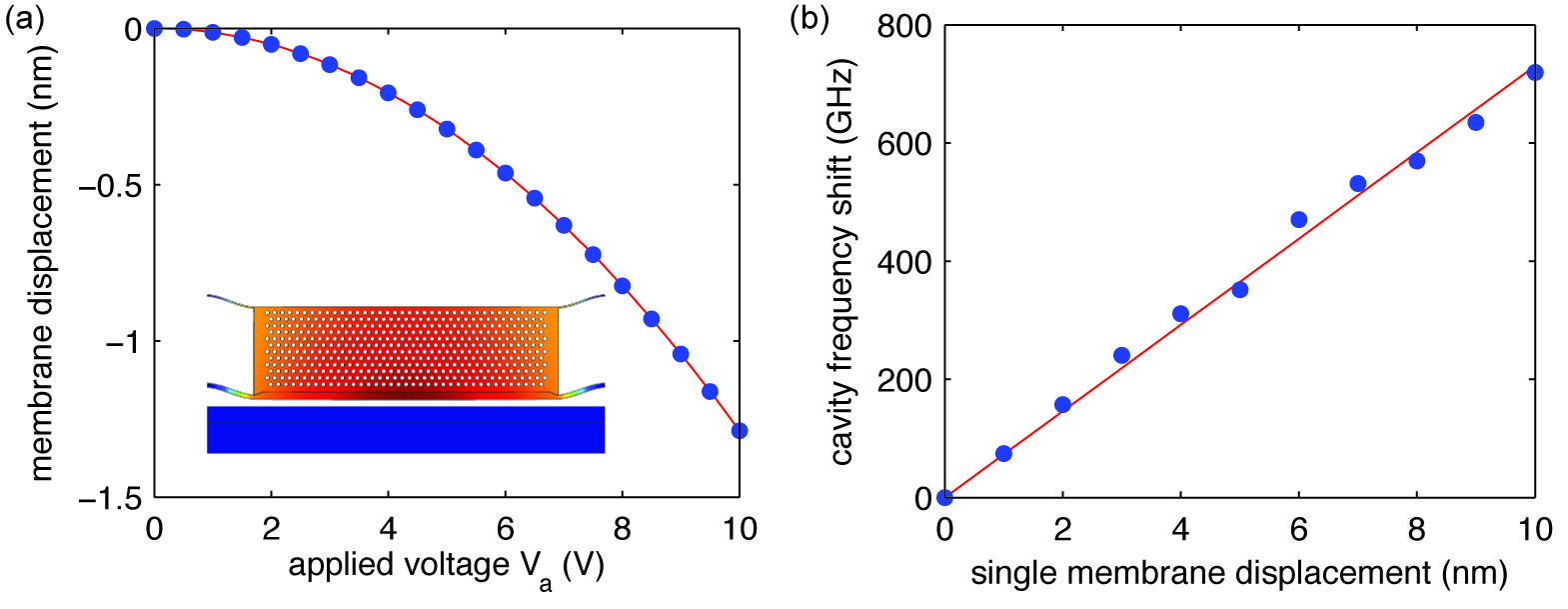}
\caption{\label{FigTuning}FEM-simulations of electrostatic tuning.
	(a) Displacement of a single cavity membrane as function of the applied voltage $V_\mathrm{a}$ for geometry parameters similar to those realized in the device studied in the main text. Blue bullets show simulation results, the red line shows a quadratic fit to the data.
	(b) Frequency shift of the fundamental cavity mode as function of the displacement of a single membrane. Within numerical error, the tuning proceeds linearly with $g'_\mathrm{OM} = 2\pi\times 73\ \mathrm{GHz/nm}$.}
\end{figure*}

The static displacement of a cavity membrane for an applied voltage $V_\mathrm{a}$ is given by balancing the capacitive force exerted by the actuator and the spring force created by a displacement of the membrane by $x$ from its rest position:
\begin{eqnarray}\label{EqBalanceForce}
F_\mathrm{cap}(x) & = & -F_\mathrm{spring}(x), \\
 \frac{1}{2}\frac{\mathrm{d}}{\mathrm{d}x}C(x)V_\mathrm{a}^2 & = & -k_\mathrm{eff}x.\label{equilibrium}
\end{eqnarray}
Here, $C(x)$ is the displacement-dependent capacitance and $k_\mathrm{eff}$ is the effective spring constant of the membrane. A realistic model for a capacitor including fringing effects is
\begin{equation}\label{capmodel}
C(x) = \frac{a}{(w_g-x)^n},
\end{equation}
where $w_g$ is the initial separation of the capacitor plates. With Eq.~(\ref{capmodel}), Eq.~(\ref{equilibrium}) has solutions for $x\leq w_g/(n+2)$. For larger displacements, the linear spring force cannot compensate the attractive capacitive force, such that the capacitor collapses \cite{Alegre10}. For the fabricated devices with $w_g = 200-250\ \mathrm{nm}$ however, such sticking occurs for displacements much larger than those realized by voltages up to 20~V (which are on the order of $10-20\ \mathrm{nm}$), such that the only limiting factor for tuning here is given by electrical breakdown of the structure. For small displacements $\delta x$, an approximate solution of Eq.~(\ref{equilibrium}) is
\begin{equation}\label{Vsquare}
\delta x = -\frac{na}{2k_\mathrm{eff}w_g^{n+1}}V_\mathrm{a}^2.
\end{equation}
For the device studied in the main text with $w_g = 230\ \mathrm{nm}$ and $w_2 = 80\ \mathrm{nm}$ we calculated the values $C\approx 0.73\ \mathrm{fF}$, $n=0.68$, and $k_\mathrm{eff} \approx 50\ \mathrm{N/m}$ using FEM simulations, such that with $a \approx Cw_g^n$ we expect $F_\mathrm{cap} \approx 1.05\ \mathrm{nN}$ and $\delta x/V_\mathrm{a}^2 \approx -21.6\ \mathrm{pm}$. A similar result is found by performing coupled electrostatics-elasticity FEM simulations that account for motion of the structure under the influence of the electrostatic force. The blue bullets in Fig.~\ref{FigTuning}(a) shows the peak membrane displacement as function of $V_\mathrm{a}$ for the studied device. Clearly, the membrane displacement follows a quadratic dependence (red line) with a displacement of $\delta x /V^2_\mathrm{a}= -13\ \mathrm{pm}$. The discrepancy between the displacement calculated in Eq.~(\ref{Vsquare}) and the result of FEM simulations arises from the capacitive force acting in a distributed manner across the entire capacitor, as opposed to the assumption of a point force made in Eq.~(\ref{EqBalanceForce}). Moreover, the intuitive model introduced above assumes one-dimensional motion of the cavity membranes. From FEM simulations, however, we find that membrane motion under the influence of capacitive actuation has a significant perpendicular component due to the different elastic moduli of silicon and gold.

In order to calculate the frequency shift of the optical cavity modes induced by the displacement of the cavity membranes, we computed the cavity resonance frequencies for a series of membrane separations using COMSOL Multiphysics \cite{COMSOL}. Results of such a simulation can be seen in Fig.~\ref{FigTuning}(b). Clearly, there is a linear relation $\Delta\omega_\mathrm{c} = g'_{i,\mathrm{OM}} x$ between displacement and frequency, with $g'_{i,\mathrm{OM}} = 2\pi\times 73\ \mathrm{GHz/nm}$ per membrane. In the presence of two actuated membranes, the total frequency shift per nanometer of membrane separation is given by $g'_\mathrm{OM} = 2g'_{i,\mathrm{OM}} = 2\pi\times 146\ \mathrm{GHz/nm}$ or a 1.1~nm wavelength shift per nanometer of displacement. The linear relation in Fig.~\ref{FigTuning}(b) persists over the entire tuning range on the order of 20~nm, consistent with the experimentally observed quadratic tuning. We note that $g'_\mathrm{OM}$ denotes the optical frequency shift induced by the electrostatically induced displacement profile which in general is not identical to a mechanical eigenmode of the structure. The optomechanical coupling constants of mechanical modes are denoted by $g_\mathrm{OM}$.

The estimated value for the voltage-dependent displacement in Eq.~(\ref{Vsquare}) together with $g'_\mathrm{OM}$ calculated above yields a predicted wavelength tunability of
\begin{equation}
\alpha = \frac{g'_\mathrm{OM}}{2\pi} \frac{\lambda_\mathrm{c}^2}{c} \frac{nC}{2k_\mathrm{eff}w_g} = 0.025\ \mathrm{nm/V^2}
\end{equation}
around $\lambda_c = 1550\ \mathrm{nm}$. From coupled electrostatics-elasticity FEM simulations we find $\alpha = 0.015\ \mathrm{nm/V^2}$. Experimentally, we observe a tunability of $0.051\ \mathrm{nm/V^2}$, approximately three times larger than the expected value. We attribute the discrepancy to the limited accuracy with which we can determine the geometrical parameters of the structure.

\section{Study of fabricated devices} \label{SecStudyFabDev}
In the design of the presented tunable cavities, the main figures to maximize tunability are the membrane stiffness expressed by $k_\mathrm{eff}$ and the capacitive force proportional to $\mathrm{d}C(x)/\mathrm{d}x$. For a plate capacitor with cross-section $A$ we have $C(w_g) = \epsilon_0 \frac{A}{w_g}$ and therefore $F_\mathrm{cap}\propto w_g^{-2}$, such that it is beneficial to minimize the capacitor gap size $w_g$. With the process used for fabrication of the metal mask, we are limited to gaps with $w_g\geq200\ \mathrm{nm}$. Further reducing the feature size could be achieved by reducing the metal layer thickness, however, this sacrifices stability in the wet chemistry process used for cavity lithography. The floppiest struts we could reproducibly fabricate had a length of $3\ \mathrm{\mu m}$ and a width of $w_2=80\ \mathrm{nm}$. The following table gives an overview of accessible tuning ranges achievable with typical device geometries.
\begin{center}
\begin{tabular}{|cl|c|}\hline
geometry		&		& tunability $\alpha$ \\ \hline
$w_g = 200\ \mathrm{nm}$	\& & $w_2 =   80\ \mathrm{nm}$	& 0.078 \\
$w_g = 200\ \mathrm{nm}$	\& & $w_2 = 130\ \mathrm{nm}$	& 0.052 \\
$w_g = 250\ \mathrm{nm}$	\& & $w_2 =   80\ \mathrm{nm}$	& 0.059 \\ \hline
\end{tabular}
\end{center}
In particular for small $w_2$, we observe a variation of $\alpha$ up to 50\% for devices processed under identical conditions. This observation justifies our explanation of the discrepancy between theoretical and experimentally observed values of $\alpha$.

\section{Analysis of mechanical spectra}\label{SecMechSpectra}
\begin{figure*}
\centering\includegraphics[width=1.8\columnwidth]{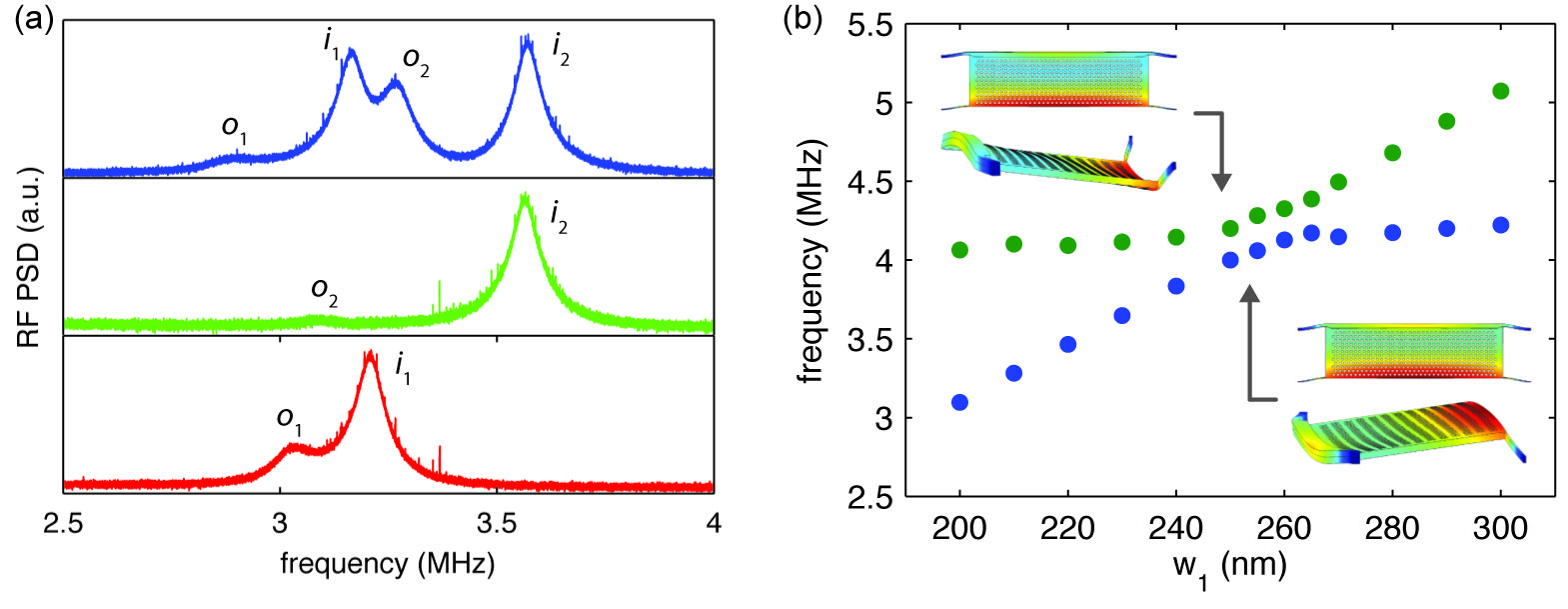}
\caption{\label{FigAntiCross1}Mechanical mode spectra as function of membrane stiffness.
	(a) Mechanical spectra for different arrangements of the fiber taper: hovering over the membranes (blue curve), touching on membrane 1 (green curve), and touching on membrane 2.
	(b) Mechanical mode frequencies as function of $w_1$, the width of the strut that carries the metal wires on the. We observe a clear anticrossing between the in-plane tuning mode and a spectrally nearby flexural mode. The device studied in the main text was processed with $w_1$ close to this anticrossing.
}
\end{figure*}

As mentioned in the main text, the complex nature of the mechanical spectra arises both from an asymmetry splitting between the two membranes as well as from a mode-mixing of in-plane and flexural modes, yielding a total of four modes. Mechanically anchoring the fiber taper individually on each membrane allows us to study the nature of the mechanical resonances. Figure~\ref{FigAntiCross1}(a) shows mechanical spectra when hovering the taper over the chip (blue line), anchoring it on membrane 1 (green curve), and anchoring it on membrane 2 (red curve). When the taper touches a certain cavity membrane, it inhibits its mechanical motion, thus removing the corresponding resonance from the spectrum. As a result, this allows us to assign the mechanical resonances to in-plane ($i_\mathrm{1},i_\mathrm{2}$) and out-of-plane ($o_\mathrm{1},o_\mathrm{2}$) motion of membranes 1 and 2.

Moreover, we find that in-plane and out-of-plane mechanical resonances are coupled due to the breaking of out-of-plane symmetry introduced by the metal wires on top of the membranes. To gain further insight into this hybridization we conducted elasticity FEM simulations of a single cavity membrane for a variety of geometry parameters. Figure~\ref{FigAntiCross1}(b) shows two mechanical eigenfrequencies as function of $w_1$, the width of the strut carrying the metal wire (see Fig.~\ref{Fig1}(e)).
Clearly, for $w_1=220\ \mathrm{nm}$, we observe an anticrossing of the modes around 4~MHz. Away from this anticrossing, the flat dispersion mode corresponds to the flexural mode shown in the left inset of Fig.~\ref{Fig2}(c), while the mode with steep dispersion has predominantly in-plane character which resembles the profile generated by capacitive actuation shown in Fig.~\ref{Fig1}(b). In this regime, DC electrostatic tuning mainly actuates the latter ``tuning'' mode. Consistent with this assumption, we calculate an optomechanical coupling constant of $g_\mathrm{OM} = 2\pi\times 152\ \mathrm{GHz/nm}$ using a perturbation Ansatz \cite{Johnson02}, in good agreement with $g'_\mathrm{OM} = 2\pi\times 146\ \mathrm{GHz/nm}$ determined in Appendix~\ref{SecTheory}. On resonance, the modes hybridize into mixed in-plane/flexural modes width displacement profiles as shown in the insets of Fig.~\ref{FigAntiCross1}(b). As a result, we expect the resonances in Fig.~\ref{FigAntiCross1}a to be mutually coupled, resulting in nontrivial frequency shifts of all modes when mechanically anchoring the taper on one of the membranes.

While the mechanical modes shown Fig.~\ref{FigAntiCross1}(a) correspond to the motion of individual membranes, the optical spring effect selectively couples to the differential motion of the two membranes. As a result, for a large optical spring, the eigenmodes are renormalized into differential (odd) and common (even) superpositions. Accordingly, the stiffening feature in Fig.~\ref{Fig3}(b) corresponds to the differential mode $i_\mathrm{1}-i_\mathrm{2}$, while the lower-frequency feature around 3.3~MHz is composed of a hybridized triplet of the even mode $i_\mathrm{1}+i_\mathrm{2}$ with the two out-of-plane modes.

The optical spring data shown in Fig.~\ref{Fig3}(b) furthermore allows us to estimate the optomechanical coupling $g_\mathrm{OM}$ of the differential in-plane mode. The normalized frequency $\omega_\mathrm{m}$ of a mechanical mode in the presence of an optical drive field is given by \begin{equation}\label{stiffening}
\omega_\mathrm{m}^2 = \omega_\mathrm{m0}^2 + \frac{2g_\mathrm{OM}^2 P_\mathrm{d}}{\omega_\mathrm{c} \kappa_i m_\mathrm{eff}} \frac{\Delta\omega}{\Delta\omega^2+\kappa^2/4},
\end{equation}
where $\Delta\omega=\omega_\mathrm{l} - \omega_\mathrm{c}$ denotes the detuning between the pump laser and the cavity mode frequency, $P_\mathrm{d} = \hbar \omega_\mathrm{c} n_\mathrm{cav} \, \kappa_i$ is the optical power dropped into the cavity, $\omega_\mathrm{m0}$ is the unperturbed circular mechanical frequency, and the intracavity photon number $n_\mathrm{cav}$ is given by
\begin{equation}
n_\mathrm{cav} = \frac{\kappa_e}{2} \frac{1}{\Delta\omega^2+\kappa^2/4} \frac{P_\mathrm{in}}{\hbar\omega_\mathrm{c}},
\end{equation}
where $P_\mathrm{in}$ is the power incident on the cavity. We extract the mode frequencies from Fig.~\ref{Fig3}(b), the detunings from the error-signal offset of the cavity-lock, and fit the data according to Eq.~(\ref{stiffening}). With $T_d = 93\%$, and the estimated $Q$ factor of $Q_i = 130\,000$ of the second order mode we arrive at $g_\mathrm{OM} = 2\pi\times 215\ \mathrm{GHz/nm}$, in reasonable agreement with the theoretical value of $2\pi\times 152\ \mathrm{GHz/nm}$.

\section{Transmission of a thermally broadened cavity mode}\label{SecVoigtFits}
The linear transmission spectrum of an optomechanically coupled cavity under the influence of thermal Brownian motion of the mechanical oscillator has been described in detail in \cite{Lin09}. In the sideband unresolved limit, the time-averaged transmission spectrum can be approximated by a Voigt profile, which describes a Lorentzian transmission line under the influence of noise that follows a Gaussian distribution. The Voigt profile is defined by the convolution integral
\begin{equation}
V(\omega) = \int_{-\infty}^{+\infty} T(\omega') G(\omega-\omega') \mathrm{d}\omega'
\end{equation}
with the Lorentzian cavity transmission function
\begin{equation}
T(\omega) = 1 - \frac{\kappa_e}{4} \frac{2\kappa-\kappa_e}{\left(\omega-\omega_\mathrm{c}\right)^2 + \kappa^2/4}
\end{equation}
and the normalized Gaussian distribution
\begin{equation}
G(\omega) = \frac{1}{\sigma \sqrt{2\pi}} e^{-\omega^2/(2\sigma^2)}.
\end{equation}
Here, $\kappa=\kappa_i + \kappa_e$ denotes the total cavity decay rate, $\kappa_i$ describes the intrinsic cavity losses, $\kappa_e$ is the total loading rate of a side-coupled double-sided cavity, and $\sigma$ is a parameter that describes Gaussian line broadening with a full width at half maximum (FWHM) of $f_G = 2\sigma\sqrt{2\ln2}$. The FWHM $f_V$ of the Voigt profile $V(\omega)$ can be approximated by \cite{Longbothum02}
\begin{equation}\label{fV}
f_V \approx 0.5346\ \kappa + \sqrt{0.2166\ \kappa^2 + f_G^2}.
\end{equation}
The magnitude of $\sigma$ can be estimated from the root-mean-square displacement of a harmonic oscillator $x_\mathrm{rms}$ under the influence of thermal Brownian motion, which according to the equipartition theorem is given by
\begin{equation}
x_\mathrm{rms} = \sqrt{\langle x^2 \rangle} = \sqrt{\frac{k_\mathrm{B}T}{m_\mathrm{eff}\omega_m^2}}.
\end{equation}
The thermally induced cavity wavelength jitter $\sigma$ then is
\begin{equation}\label{sigma}
\sigma = \frac{g_\mathrm{OM}}{2\pi} \frac{\lambda_\mathrm{c}^2}{c} x_\mathrm{rms}.
\end{equation}
For the calculated system parameters $g_\mathrm{OM} = 2\pi\times 152\ \mathrm{GHz/nm}$, $m_\mathrm{eff} = 2\cdot72\ \mathrm{pg}$, and with $Q_i = 200\,000$ and $\omega_\mathrm{m} = 2\pi\times 3.6\ \mathrm{MHz}$ extracted from experiments, combining eqs.~(\ref{fV}) \& (\ref{sigma}) yields 24.7~pm, in good agreement with the experimentally observed FWHM of 22~pm.

\section{AC-tuning of the structure}\label{SecAcTuning}
As mentioned in the main text, we can resonantly address the mechanical modes of the structure by applying an AC voltage to the cavity contacts. The use of metallic contacts on the cavities is beneficial to this end, as it provides for a serial resistance on the order of several $10\ \Omega$, estimated from closed-loop structures of similar dimensions. Together with the total device capacitance on the order of several $10\ \mathrm{fF}$, this yields an RC cut-off orders of magnitude above the mechanical mode frequencies. Applying a sinusoidal drive field $V(t) = V_\mathrm{AC} \cos(\omega t)$ to the contacts generates a force
\begin{eqnarray}
F_\mathrm{cap}(t) & = & \frac{1}{2} \frac{\mathrm{d}C}{\mathrm{d}x} V^2_\mathrm{AC} \cos^2(\omega t) \nonumber \\
 & = & F_\mathrm{AC} \left[ 1+\cos(2\omega t) \right]
\end{eqnarray}
acting on the membrane, with
\begin{equation}
F_\mathrm{AC} = \frac{1}{2} \frac{\mathrm{d}C}{\mathrm{d}x} \left(\frac{V_\mathrm{AC}}{\sqrt{2}} \right)^2.
\end{equation}
$F_\mathrm{cap}$ comprises a DC-part and an oscillating component at angular frequency $2\omega$, which creates oscillations of the membranes according to
\begin{equation}
x(t) = \frac{F_\mathrm{AC}}{m_\mathrm{eff}} \frac{1}{\sqrt{(4\omega^2-\omega_i^2)^2 + 4\omega^2\gamma_\mathrm{m}^2}} \cos(2\omega t)
\end{equation}
in the presence of a mechanical mode at $\omega_i$. When the double drive frequency $2\omega$ is resonant with a mechanical mode frequency $\omega_i$, this yields a modulation of the cavity resonance with amplitude
\begin{equation}\label{EqAcTuning}
\Delta\omega_{c,\mathrm{AC}} = g_\mathrm{OM} \frac{F_\mathrm{AC}}{k_{i,\mathrm{eff}}} Q_{m,i}
\end{equation}
with $k_{i,\mathrm{eff}} = m_\mathrm{eff}\omega_i^2$ the effective spring constant of mode $i$, and $Q_{m,i} = \omega_i/\gamma_\mathrm{m}$ its $Q$-factor.

\begin{figure}[ht!]
\centering\includegraphics[width=1\columnwidth]{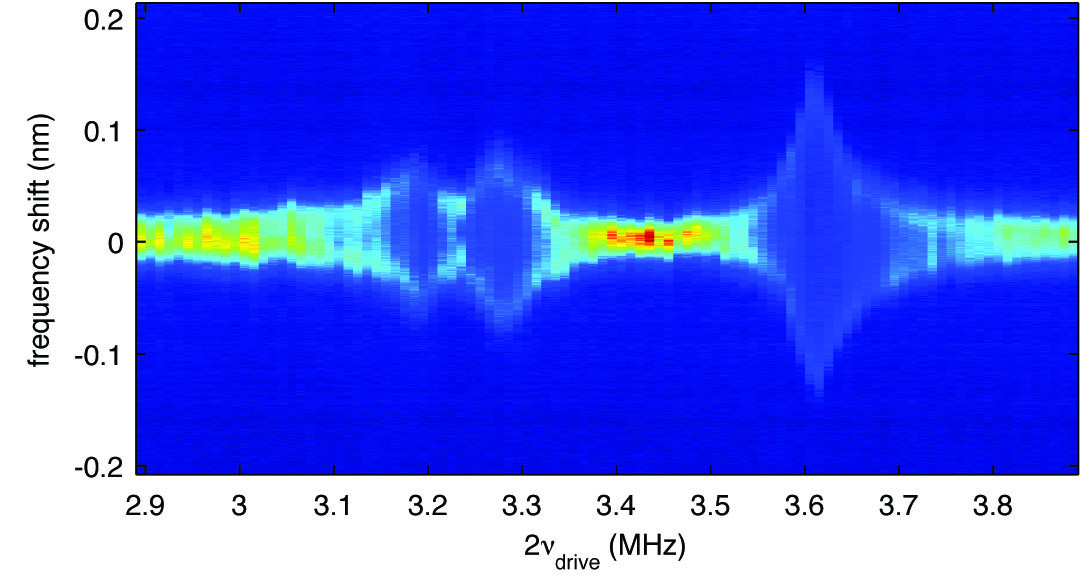}
\caption{\label{ACtuning}AC-tuning curve of the device used for Fig.~\ref{Fig2}(c) of the main text. The false-color plot shows transmission spectra of the fundamental cavity mode when applying a sinusoidal drive voltage to the cavity contacts. Mechanical resonances are clearly visible by broadening of the transmission feature, when the double drive frequency $2\nu$ coincides with a mode frequency. The mechanical mode structure found form RF PSD measurements is clearly reproduced here.}
\end{figure}

Figure~\ref{ACtuning} shows a series of transmission spectra acquired on the fundamental cavity mode in a false-color plot, while applying an AC voltage with an amplitude of 200~mV to the contacts and sweeping the drive frequency $\nu_\mathrm{drive}$. At $2\nu_\mathrm{drive} = $3.18~MHz, 3.28~MHz, and 3.61~MHz, we observe significant broadening of the transmission spectrum. These frequencies correspond to the mechanical mode frequencies observed in Fig.~\ref{Fig2}(c). Resonant electromechanical actuation of the modes results in fast oscillation of the optical cavity resonance according to Eq.~(\ref{EqAcTuning}). Since the spectra in Fig.~\ref{ACtuning} are acquired on timescales much longer than a mechanical oscillation period, we observe an averaged line-shape, resulting in the broadening features observed in Fig.~\ref{ACtuning}.

This results renders resonant electromechanical actuation of mechanical modes a viable tool for the spectroscopy of the mechanical mode structure.

\section{Study of optomechanical cooling and lasing}\label{SecAppBackaction}
The presence of mode-mixing described in Appendix~\ref{SecMechSpectra} makes optomechanical effects like phonon lasing and back-action cooling difficult to interpret. In order to single out an individual resonance, we anchored the fiber taper on one of the membranes, thus blocking its mechanical motion and leading to a mechanical RF-spectrum like the one shown in the centre tab of Fig.~\ref{FigAntiCross1}(a). The increasing mechanical linewidth of the resonance shown in Fig.~\ref{Fig4}(d) is indicative of optomechanical cooling, as is the reduction of the area under the peak. Surprisingly, however, the data shown in Fig.~\ref{Fig4}(e) does not exactly follow a linear dependence. We argue that the reason for this is twofold: firstly, the observed mechanical resonance is still subject to mode mixing with other modes. Moreover, thermal motion of the cavity membrane is expected to lead to an averaging effect of optomechanical cooling. The damping rate of a mechanical resonance under the influence of optomechanical back-action can be written as
\begin{equation}\label{EqLinewidth}
\gamma_\mathrm{m} = \gamma_{\mathrm{m},0} - \frac{2g_\mathrm{OM}^2 P_\mathrm{d} \kappa}{\omega_\mathrm{c} \kappa_i m_\mathrm{eff}} \frac{\Delta\omega}{(\Delta\omega^2+\kappa^2/4)^2}.
\end{equation}
While in the experiment we actively lock the cavity to a detuning of $\Delta\omega = -\kappa/2$ from the pump laser, thermal motion of the cavity membrane leads to fluctuations in $\Delta\omega$. Since the dynamics of the optical cavity mode is much faster than the timescales over which mechanical fluctuations occur (our device is deep in the sideband-unresolved regime: $\kappa \gg \omega_\mathrm{m}$), this leads to an averaging of $\gamma_\mathrm{m}$. With increasing pump power, the temperature and thereby the motional amplitude of the mechanical mode is reduced, thus explaining the roll-off of the data for small $n_\mathrm{cav}$. A phenomenological model in which the result of R|Eq.~(\ref{EqLinewidth}) is convolved with a Gaussian frequency distribution and that assumes realistic parameters qualitatively confirms this interpretation. Moreover, the magnitude of cooling observed in Fig.~\ref{Fig4}(e) is consistent with $g_\mathrm{OM} \approx 2\pi \times 150\ \mathrm{GHz/nm}$.

Finally, we check for quantitative consistency of the lasing threshold observed in the data shown in Fig.~\ref{Fig4}(b). Experimentally, we drive the system with a constant incident power of $P_\mathrm{i} = 250\ \mathrm{\mu W}$ and sweep $\Delta\omega$ using electrostatic tuning in open loop configuration. Experimentally, we find the lasing threshold to occur for a detuning of $\Delta\omega \approx +0.65\kappa$, where the intracavity photon number is estimated to be $n_\mathrm{cav} \approx 13\,000$. It has to be noted that determination of the experimentally realized detuning $\Delta\omega$ is difficult due to the bistability of the cavity transmission curve that arises from time-averaging of the pump laser absorption when the mechanical mode is strongly driven in the lasing regime. From the observed values we extract $g_\mathrm{OM} = 2\pi \times 93\ \mathrm{GHz/nm}$, in reasonable agreement with the observations made above.

\end{document}